\documentclass[%
prl,
superscriptaddress,    
twocolumn,showpacs,floats,aps
]{revtex4-1}

\usepackage{graphicx}

\usepackage{wasysym}
\usepackage{color}
\usepackage{amsfonts,amssymb}

\usepackage{rotating}

\def\ket#1{|#1\rangle}
\def\bra#1{\langle#1|}
\def\braket#1#2{\langle#1|#2\rangle}
\newcommand{\comment}[1]{}

\begin{document}

\title{Hybrid NRG-DMRG approach to real-time dynamics of
       quantum impurity systems}

\author{Fabian G\"uttge}
\author{Frithjof B.~Anders}
\affiliation{Lehrstuhl f\"ur Theoretische Physik II,
             Technische Universit\"at Dortmund,
             44221 Dortmund,Germany}

\author{Ulrich Schollw\"ock}
\affiliation{Physics Department, Arnold Sommerfeld Center
             for Theoretical Physics, and Center for
             NanoScience, Ludwig-Maximilians-Universit\"at
             M\"unchen, D-80333 M\"unchen, Germany}

\author{Eitan Eidelstein}
\affiliation{Racah Institute of Physics, The Hebrew
             University, Jerusalem 91904, Israel}
\affiliation{Department of Physics, NRCN, P.O. Box
             9001, Beer-Sheva, 84190 Israel}

\author{Avraham Schiller}
\affiliation{Racah Institute of Physics, The Hebrew
             University, Jerusalem 91904, Israel}

\begin{abstract}
A hybrid approach to nonequilibrium dynamics of
quantum impurity systems is presented. The numerical
renormalization group serves as a means to generate
a suitable low-energy Hamiltonian, allowing for an
accurate evaluation of the real-time dynamics of the
problem up to exponentially long times using primarily
the time-adaptive density-matrix renormalization group.
We extract the decay time of the interaction-enhanced
oscillations in the interacting resonant-level model
and show their quadratic divergence with the interaction
strength $U$. Our numerical analysis is in excellent
agreement with analytic predictions based on an
expansion in $1/U$.
\end{abstract}

\maketitle


\paragraph{Introduction.}

The description of strong electronic correlations
far from thermal equilibrium poses an enormous
theoretical challenge. At the root of the problem
lies the nonequilibrium density operator which is
not explicitly known in the presence of interactions.
Of particular relevance are quench and driven dynamics
realized in pump-probe
experiments~\cite{Spectroscopy-1,Spectroscopy-2},
atomic traps~\cite{cold-atoms,cold-atoms-3},
and nanodevices~\cite{Elzerman04,Petta05},
where the full time evolution of the density
operator should, in principle, be tracked. 

Quantum impurities systems (QIS) have regained
considerable attention over the past 15 years due to
the advent of carefully designed nanodevices. These
generically consist of a few locally interacting
degrees of freedom, typically a quantum dot, in
contact with macroscopic leads. Since driven dynamics
in nanodevices is of practical relevance to quantum
computing and quantum control, considerable efforts
were mounted in recent years toward devising approaches
capable of treating the nonequilibrium state in QIS.

Significant analytical progress in the calculation
of real-time dynamics was achieved using
different adaptations of perturbative
renormalization-group ideas~\cite{RT-RG,
Flow-equations, FRG}. However, with the 
exception of Ref.~\cite{RT-RG-Strong-coupling},
these are confined to the weak-coupling regime.
Numerical methods, such as applications of
the time-dependent density-matrix renormalization
group (TD-DMRG)~\cite{TD-DMRG,Schollwock2011} to
QIS~\cite{Schmitteckert,Feiguin},
the time-dependent numerical renormalization group
(TD-NRG)~\cite{TD-NRG}, an iterated path-integral
approach~\cite{Weiss-et-al-2008}, and different
continuous-time Monte Carlo
simulations~\cite{Muehlbacher-rabani-2008,
Werner-et-al,Schiro-Fabrizio-2009},
are more flexible in the parameter regimes they
can treat, but are either restricted to short time
scales~\cite{Schmitteckert,Feiguin,Weiss-et-al-2008,
Muehlbacher-rabani-2008,Werner-et-al,Schiro-Fabrizio-2009}
or susceptible to finite-size and discretization
errors~\cite{Schmitteckert,Feiguin, TD-NRG}. Indeed,
finite-size representations are faced with an inherent
difficulty of accurately representing the continuum
limit even on intermediate time scales.

In this paper, we report the extension of a recent
hybrid approach~\cite{Hybrid-NRG} that overcomes some
of the major obstacles hampering the description of
quench dynamics in QIS. The basic idea is to
exploit the outstanding capabilities of the TD-NRG
to bridge over vastly different time scales in order
to systematically construct an effective low-energy
Hamiltonian, whose real-time dynamics can be calculated
using complementary approaches that do not rely on
the special structure of the Wilson chain. In this
manner, one can largely eliminate discretization
errors inherent to the Wilson chain while boosting
the complementary approach to times scales orders
of magnitude beyond its natural capabilities.
As a proof of principle, we have hybridized in
Ref.~\cite{Hybrid-NRG} the TD-NRG with the Chebyshev
expansion technique (CET)~\cite{TalEzer-Kosloff-84}.
In this paper, we demonstrate the full power of the
approach by hybridizing the TD-NRG with the
TD-DMRG~\cite{Schollwock2011}.

Focusing on the interacting resonant-level model
(IRLM)~\cite{VigmanFinkelstein78,Schlottmann80},
we show that one can essentially eliminate
discretization errors on all time scales of
interest by constructing a suitable hybrid chain.
This, in turn, allows for a thorough examination  of
the interaction-enhanced oscillations first
reported in Ref.~\cite{Hybrid-NRG}, yielding
excellent agreement with analytical predictions for
their frequency and damping time. The latter is shown
to diverge quadratically with the interaction strength,
demonstrating that relaxation to equilibrium can
involve new time scales far longer than the
thermodynamic ones.

\paragraph{Hybrid-NRG.}

We begin with a concise derivation of the
hybrid-NRG~\cite{Hybrid-NRG}. The Hamiltonian
${\cal H} = {\cal H}_{\rm bath} + {\cal H}_{\rm imp}
+ {\cal H}_{\rm mix}$ of a quantum impurity problem
consists of three parts: ${\cal H}_{\rm bath}$ models
the continuous bath, ${\cal H}_{\rm imp}$ represents the
decoupled impurity, and ${\cal H}_{\rm mix}$ describes
the coupling between the two subsystems. For $t < 0$,
the entire system is assumed to be characterized by a
density operator $\hat{\rho}_0$ associated with an
initial Hamiltonian ${\cal H}^i$. Specifically,
$\hat{\rho}_0$ can either be the equilibrium density
operator corresponding to ${\cal H}^i$, or may project
onto one of its low-lying eigenstates, typically the
ground state. At time $t = 0$, a static perturbation
is abruptly switched on such that
${\cal H}^i \to {\cal H}^f$. Our goal is to track
the time evolution of local expectation values:
$O(t) = {\rm Tr} \{ \hat{\rho}(t) \hat{O} \}$ with
$\hat{\rho}(t) = e^{-i t {\cal H}^f} \hat{\rho}_0
e^{i t {\cal H}^f}$. 

In Wilson's numerical renormalization group
(NRG)~\cite{NRG}, ${\cal H}_{\rm bath}$ is discretized
logarithmically using a dimensionless parameter
$\Lambda > 1$, and mapped onto a semi-infinite
chain whose open end is coupled to the impurity via
${\cal H}_{\rm mix}$. Wilson's chain is characterized
by exponentially decreasing hopping matrix elements
$t_m \propto D \Lambda^{-m/2}$, defining a natural
separation of scales. This enables an iterative
diagonalization of ${\cal H}$, where at each step
only the lowest $N_s$ eigenstates are retained.
Terminating the procedure after $N$ steps, the
collection of states discarded after each iteration
combine to form a complete basis set of approximate
NRG eigenstates of ${\cal H}$ on the $N$-site
chain~\cite{TD-NRG}.
The expectation value of any local operator $\hat{O}$
can be formally expressed as~\cite{TD-NRG}
\begin{equation}
O(t \ge 0) =
        \sum_{m=0}^{N}
             \sum_{r,s}^{\rm trun} \;
                  O_{r,s}^m \rho^{m}_{s,r}(t) ,
\label{eqn:time-evolution-intro} 
\end{equation}
where $r$ and $s$ run over the NRG eigenstates
of ${\cal H}^f $ at iteration $m \le N$,
$O_{r,s}^m$ is the matrix representation of
$\hat O$ at that iteration, and $\rho^{m}_{s,r}(t)$
is the corresponding time-dependent reduced density
matrix. The restricted sum over $r$ and $s$ requires
that at least one of these states is discarded at
iteration $m$.

Partitioning the sum over $m$ into $m \leq M$ and
$M < m$ at some arbitrary but fixed $M < N$,
Eq.~(\ref{eqn:time-evolution-intro}) is recast as
\begin{equation}
O(t \geq 0) = \sum_{m=0}^{M}
              \sum_{r,s}^{\rm trun} \;
                   O_{r,s}^m \rho^{m}_{s,r}(t) 
              + {\rm Tr}
                     \{
                        \hat 1_M^+ \hat O \hat 1_M^+
                        \hat \rho(t) \hat 1_M^+
                     \} ,
\label{eqn:time-evolution-hybrid} 
\end{equation}
where $\hat 1_M^+$ projects onto the subspace retained
at the conclusion of iteration $M$.
Equation~(\ref{eqn:time-evolution-hybrid}) is formally
exact, relying solely on the completeness of our basis
set~\cite{TD-NRG}. It has the following interpretation.
At each energy scale $D \Lambda^{-m/2}$ with
$m \leq M$, only those terms involving at least
one discarded high-energy state contribute to $O(t)$,
leaving the contribution of the low-energy subspace
retained at the conclusion of iteration $M$. In the
process, the NRG has produced an effective
quantum-impurity Hamiltonian ${\cal H}_{M + 1}$ with the
reduced bandwidth $D_{\rm eff} \propto D \Lambda^{-M/2}$:
\begin{eqnarray}
{\cal H}_{M + 1} &=&
         \sum_{k} E^{M}_{k} \ket{k;M}\bra{k;M}
\nonumber \\
       &+& \sum_{m = M}^{N - 1}
           \sum_{\nu}
                t_{m}
                \hat 1_M^+
                \{ f^\dagger_{m+1, \nu} f_{m, \nu}
                   + {\rm H.c.} \}
                \hat 1_M^+ .
\end{eqnarray}
Here, $f^{\dagger}_{m, \nu}$ creates an electron of
flavor (spin) $\nu$ on the chain site $m$, $\ket{k;M}$
labels the kept NRG eigenstates at iteration $M$,
and $E_{k}^M$ are the corresponding NRG eigenenergies.
Usually, one would proceed with the NRG to
iteratively diagonalize ${\cal H}_{M + 1}$.
Here, we follow a different route:
(i) we abandon the traditional Wilson chain and
    seek an optimal choice for the hopping amplitudes
    $t_{m}$ with $m \geq M$;
(ii) ${\cal H}_{M + 1}$ is used as input for our
     complementary method of choice in order to compute
     ${\rm Tr} \{ \hat 1_M^+ \hat O \hat 1_M^+
     \hat{\rho}(t) \hat 1_M^+ \}$;
(iii) employing the standard NRG approximation
      $\rho_{s, r}^{m}(t) \approx e^{i (E_r^m - E_s^m)t}
      \rho^{\rm red}_{s, r}(m)$~\cite{TD-NRG},
      $\rho^{\rm red}_{s,r}(M)$ as produced by our
      method of choice is feed back into the TD-NRG to
      account for the remaining high-energy dynamics
      in Eq.~(\ref{eqn:time-evolution-hybrid}).

In this paper, we supplement the TD-NRG with the
adaptive TD-DMRG~\cite{Schollwock2011}. The system
is assumed to initially occupy the ground state
$\ket{\psi_0}$ of ${\cal H}_{M + 1}^{i}$,
constructed using the DMRG~\cite{DMRG}.
Accordingly, $\hat{\rho}(t)$ equals
$\ket{\psi(t)}\bra{\psi(t)}$ with
$\ket{\psi(t)} = e^{-i t {\cal H}^f} \ket{\psi_0}$,
resulting in ${\rm Tr} \{ \hat 1_M^+ \hat O
\hat 1_M^+ \hat \rho(t)\hat 1_M^+ \} =
\bra{\chi_M(t)}\hat O\ket{\chi_M(t)}$. Here
$\ket{\chi_M(t)} = \hat 1_M^+ \ket{\psi(t)}$ is
the projection of $\ket{\psi(t)}$ onto the low-energy
subspace retained at the conclusion of iteration $M$. 
Although Eq.~(\ref{eqn:time-evolution-hybrid}) is
formally exact for arbitrary $M$, the larger is
$N_\chi = \braket{\chi_M(t)}{\chi_M(t)} \le 1$ the
smaller is the contribution of the sum on the right-hand
side of Eq.~(\ref{eqn:time-evolution-hybrid}). If
$(1 - N_\chi) < \epsilon$ for some small number
$\epsilon$, then $\ket{\chi_M(t)} = \ket{\psi(t)}
+ {\cal O}(\sqrt{\epsilon})$,
and the major contribution to the real-time dynamics
originates from $\bra{\chi_M(t)} \hat O \ket{\chi_M(t)}$.

A proper choice of $M$ is important. Initially,
the NRG level flows~\cite{NRG} of ${\cal H}^i$ and
${\cal H}^f$ are nearly identical. We choose $M$ to
be a characteristic iteration after which the two
level flows begin to significantly deviate from one
another~\cite{TD-NRG,NRG}. The corresponding energy
scale, $D \Lambda^{-M/2}$, is typically of order the
energy difference between ${\cal H}^i$ and
${\cal H}^f$. By that choice of
$M$, the major contribution to $O(t)$ stems from 
$\bra{\chi_M(t)}\hat O\ket{\chi_M(t)}$. We approximate 
$\ket{\chi_M(t)}$ by $\exp ( -i {\cal H}^{f}_{M + 1} t )
\hat 1_M^+ \ket{\psi_0}$,  adopting the NRG philosophy
that excitations on different energy scales are only
weakly coupled~\cite{Hybrid-NRG}. Thus, the NRG
generates a suitable low-energy Hamiltonian, allowing
for the real-time dynamics of the problem to be explored
on the exponentially long time scale $1/D_{\rm eff}
\propto \Lambda^{M/2}$ using mainly the TD-DMRG.

\paragraph{The model.}

We shall demonstrate our hybrid NRG-DMRG approach
by investigating the quench dynamics in the IRLM,
defined by the Hamiltonian
\begin{eqnarray}
{\cal H} &=&
         \sum_{k}
              \epsilon_{k} c^{\dagger}_{k} c^{}_{k}
         + E_d d^{\dagger} d
         + \frac{V}{\sqrt{N_k}}
           \sum_{k}
              \bigl \{
                       c^{\dagger}_{k} d
                       + {\rm H.c.}
              \bigr \}
\nonumber\\
        &+& \frac{U}{N_k}
            \sum_{k, k'}
                  :\! c^{\dagger}_{k} c^{}_{k'} \!: 
           \left (
                   d^{\dagger} d - \frac{1}{2}
           \right ) .
\label{eqn:IRLM}   
\end{eqnarray}
Here, $d^{\dagger}$ creates an electron on the
impurity level, $c^{\dagger}_{k}$ creates a band
electron with momentum $k$, $N_k$ denotes the
number of distinct $k$ points, and
$:\! c^{\dagger}_{k} c^{}_{k'} \!\!: =
c^{\dagger}_{k} c^{}_{k'} - \delta_{k, k'}
\theta(-\epsilon_{k})$ stands for normal ordering
with respect to the filled Fermi sea. The basic
energy scales in the problem include the level
energy $E_d$ and the hybridization width
$\Gamma_{0} = \pi \varrho V^2$, where $\varrho$
is the conduction-electron density of states at
the Fermi energy. The effect of the contact
interaction is to renormalize the hybridization
width at resonance according to
$\Gamma_0 \to \Gamma_{\rm eff} \approx
D (\Gamma_0/D)^{1/(1 + \alpha)}$, with
$\alpha = 2 \delta - \delta^2$ and
$\delta = (2/\pi) \arctan(\pi \varrho U/2)$
($D$ is the bandwidth). Although the thermodynamics
of the IRLM was investigated over 30 years
ago~\cite{VigmanFinkelstein78,Schlottmann80},
there has been a recent surge of interest in its
nonequilibrium properties, particularly for a
biased two-lead setting~\cite{RT-RG,FRG,MethaAndrei2005,
BoulatSaleurSchmitteckert2008}. Focusing on the
single-band version of Eq.~(\ref{eqn:IRLM}), we
consider an abrupt shift of the level energy at time
$t = 0$ from $E_d^{i}$ to $E_d^{f}$, with the goal of
tracking the time evolution of the level occupancy,
$n_d(t) = \langle d^{\dagger}(t)d (t) \rangle$.

\begin{figure}[tb]
\centering
\includegraphics[width=0.48\textwidth]{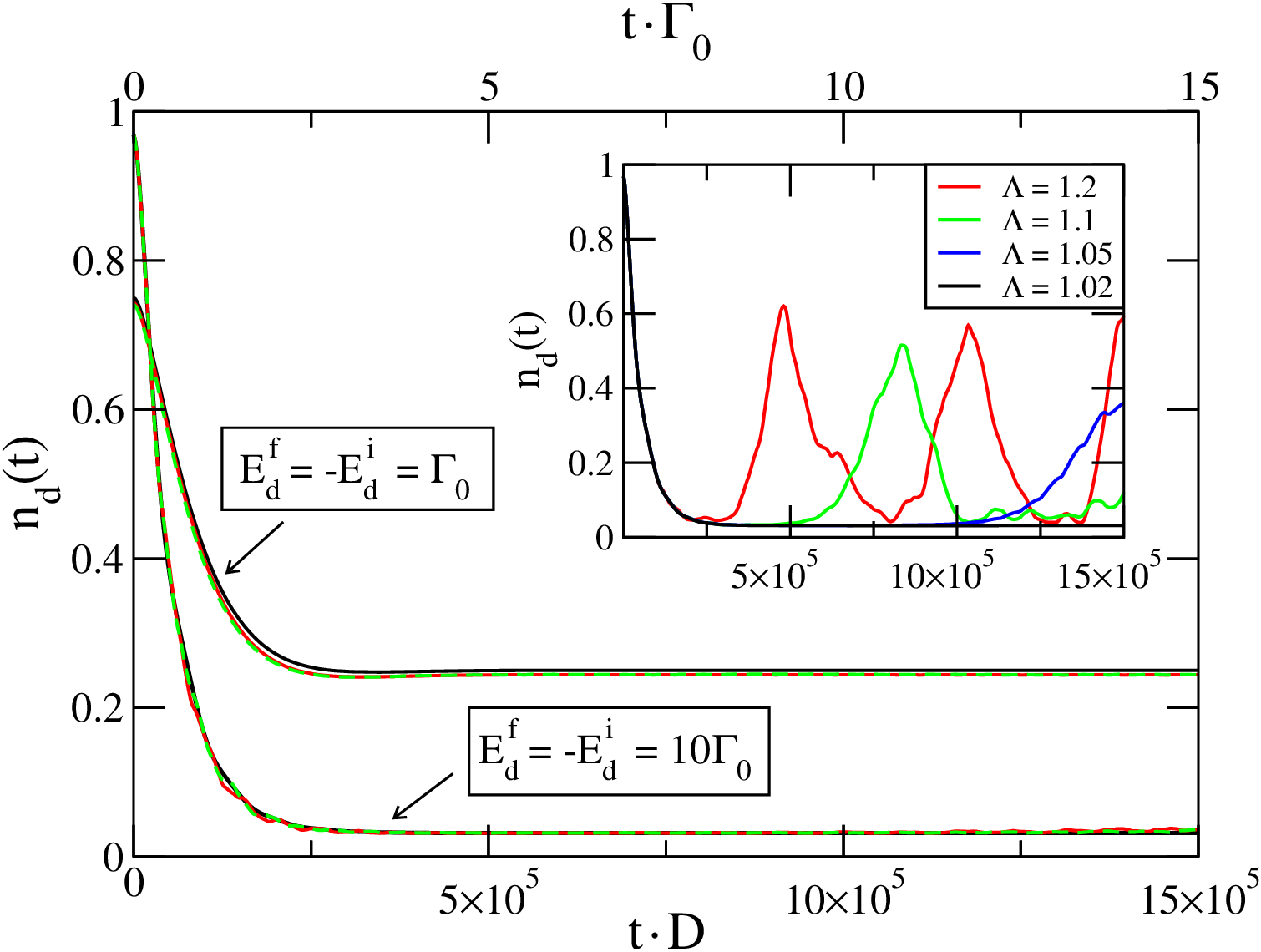}
\caption{(Color online)
         Time evolution of $n_d(t)$ on a double
         Wilson chain, following a sudden change
         of the level energy from $E_d^{i}$ to
         $E_d^{f}$. Here $\Gamma_0/D = 10^{-5}$
         and $U = 0$.
         The full red line depicts the exact solution
         on the double Wilson chain (obtained by exact
         diagonalization), the dashed green line
         displays the hybrid NRG-DMRG, and the full
         black line is the exact analytical continuum-limit
         solution~\cite{TD-NRG} in the wide-band limit.
         Chain parameters: $M = 29$, $\Lambda_1 = 1.8$,
         $\Lambda_2 = 1.02$, $N_{\rm inter} = 4$, and
         $N = 180$. $N_s = 50$ states are retained both
         in the NRG and in the course of the TD-DMRG.
         Inset: Exact $n_d(t)$ on a pure Wilson chain,
         for $E_d^{f} = -E_d^{i} = 10 \Gamma_0$ and 
         different $\Lambda$'s.
         Here $N = 500$ ($1000$) for $\Lambda \geq 1.1$
         ($\Lambda \leq 1.05$).
        }
\label{fig:Fig1}
\end{figure}

\paragraph{Hybrid chain.}

There are two sources of deviations from the
continuum limit when considering quench dynamics
on a pure Wilson chain:
(i) internal reflections of currents caused by the
exponentially decreasing hopping matrix elements
along the chain~\cite{Hybrid-NRG,Schmitteckert2010}
(leading, in turn, to an exponential slowing down
of the transport velocity);
(ii) reflections at the end of the finite-size
chain that propagate back to the impurity. While the
former source of error is eliminated for an ordinary
tight-binding chain, the latter point is unavoidable
in nearly all practical calculations as the total
chain length is limited by computational demands.
Thus, one would like to simultaneously minimize the
internal reflections and the transport velocity
down the chain to accurately access long times.

Guided by these considerations, we found it
advantageous to use a double Wilson chain, constructed
by patching two separate Wilson chains. The first $M$
hopping matrix elements $t_m$ with $0 \leq m \leq M -1$
are taken to be the customary Wilson hopping
amplitudes~\cite{NRG} with the discretization
parameter $\Lambda_1$. Further down the chain a
second, smaller discretization parameter $\Lambda_2$
is used, with a magnitude close to but larger than
one~\cite{Note1}. To reduce internal reflections,
the transition from $\Lambda_1$ to $\Lambda_2$ is
smoothed according to
$t_{M + m} = \lambda_m^{-1/2} t_{M + m - 1}$ with
\begin{equation}
\lambda_m =
\left \{
         \begin{array}{ll}
               \Lambda_1 - \frac{\Lambda_1 - \Lambda_2}
                                {N_{\rm iter}} (m + 1) ,
               & 0 \leq m < N_{\rm inter} ,
               \\
               \Lambda_2 , & N_{\rm iter} \leq m .
        \end{array}
\right .
\end{equation}

\paragraph{Results.}

The merit of such a double Wilson chain is demonstrated
in Fig.~\ref{fig:Fig1}, where $n_d(t)$ is plotted
following a quench from $E_d^{i}$ to $E_d^{f}$. The
interaction $U$ is set to zero, to facilitate
comparison with an exact continuum-limit
solution in the wide-band limit~\cite{TD-NRG}, as
well as with an exact numerical solution on the
hybrid chain, obtained by exact diagonalization
of the single-particle eigenmodes. We set
$\Gamma_0/D = 10^{-5}$, placing the basic time scale
$1/\Gamma_0$ orders of magnitude beyond the reach
of pure TD-DMRG. The parameter $M$ was chosen such
that $D \Lambda_1^{-M/2} \approx 20\Gamma_0$ is
twice the maximal value of $|E_d|$ used. 

\begin{figure}[tb]
\centering
\includegraphics[width=0.48\textwidth]{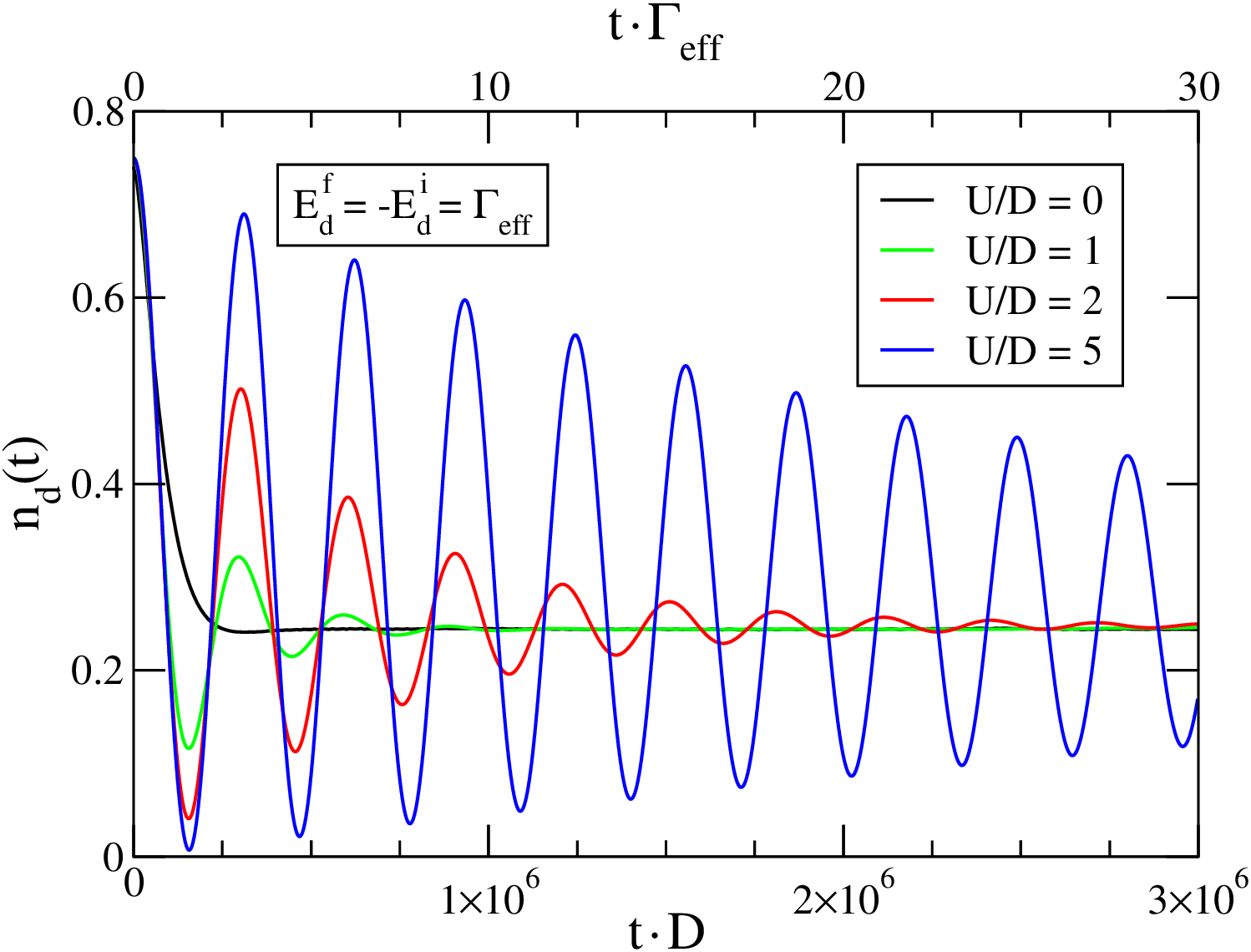}
\caption{(Color online)
         Same as Fig.~\ref{fig:Fig1}, for
         $E_d^{f} = -E_d^{i} = \Gamma_{\rm eff}$
         and different values of $U$. $\Gamma_0$
         was adjusted separately for each value
         of $U$ so as to maintain $\Gamma_{\rm eff}/D
         = 10^{-5}$. All other chain, NRG, and
         TD-DMRG parameters are the same as in
         Fig.~\ref{fig:Fig1}.
        } 
\label{fig:Fig2}
\end{figure}

Evidently, deviations between the continuum
limit, the exact solution on the hybrid chain, and
the hybrid NRG-DMRG approach are hardly discernible
up to long time scales, well after the occupancy
has relaxed to its new equilibrium value. For
$E_d^{f} = -E_d^{i} = \Gamma_0$, the excellent
agreement persists up to $t \agt 30/\Gamma_0$,
at which point all three curves begin to separate.
For $E_d^{f} = -E_d^{i} = 10\Gamma_0$, the agreement
extends up to slightly above $15/\Gamma_0$. As
analyzed in the inset, an impractically small
discretization parameter $\Lambda \approx
\Lambda_2 = 1.02$ is needed to achieve a
comparable representation of the continuum limit
using a pure Wilson chain.

In Fig.~\ref{fig:Fig2}, we analyze the effect of
a finite $U$ on $n_d(t)$. At low energies, the IRLM
is equivalent to its noninteracting counterpart,
both describing a phase-shifted Fermi liquid. Near
resonance, the effect of $U$ in equilibrium is to
renormalize $\Gamma_0$ to $\Gamma_{\rm eff}$,
hence one may expect $n_d(t)$ to follow the same
curves as in Fig.~\ref{fig:Fig1} upon substituting
$\Gamma_0 \to \Gamma_{\rm eff}$. This, however,
is not the case. In Fig.~\ref{fig:Fig2} we adjusted
$\Gamma_0$ separately for each value of $U$ so as
to maintain a fixed $\Gamma_{\rm eff}/D = 10^{-5}$.
To this end, we fixed $E_d/D = -10^{-5}$ and scanned
$\Gamma_0$ using the hybrid NRG-DMRG until a
ground-state occupancy of $n_d = 0.75$ was reached.
As first reported in Ref.~\cite{Hybrid-NRG},
interaction-enhanced oscillations gradually
develop in $n_d(t)$ upon increasing $U$.
For large $U$, exemplified by $U/D = 5$,
these oscillations decay on a time scale much longer
than $1/\Gamma_{\rm eff}$, revealing the emergence of
a new time scale unrelated to the thermodynamic ones.
Note that, as for $U = 0$, our curves appear
to faithfully represent the continuum limit up
to $t \agt 30/\Gamma_{\rm eff}$ for this  moderate
quench.

\begin{figure}[tb]
\centering
\includegraphics[width=0.48\textwidth]{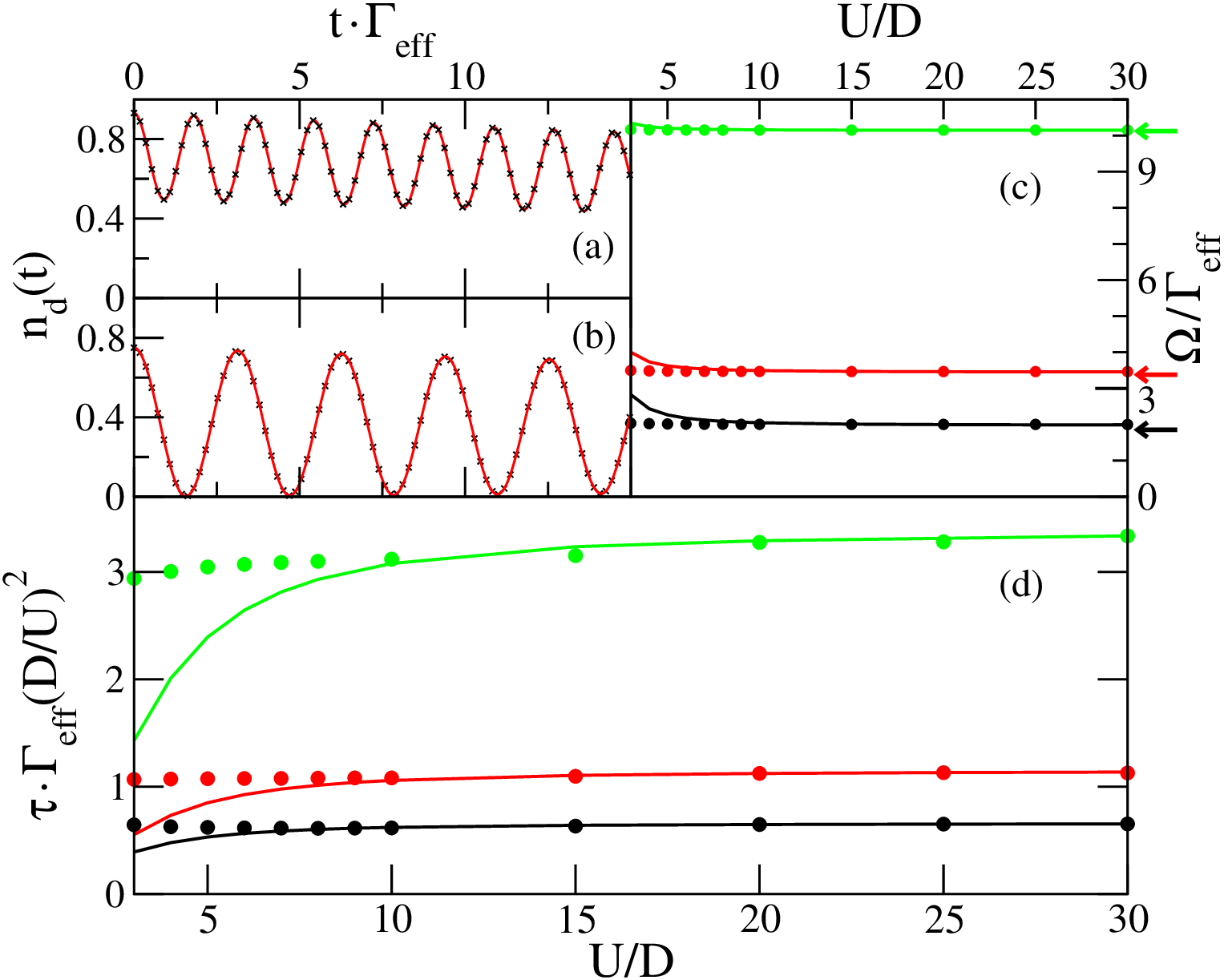}
\caption{(Color online)
         (a) Interaction-enhanced oscillations (red
             line) for $E_d^{f} = -E_d^{i} =
             3\Gamma_{\rm eff}$ and $U/D = 10$, along 
             with a fit to Eq.~(\ref{fit-function})
             (black crosses).
         (b) Same as (a), for $E_d^{f} = -E_d^{i} =
             \Gamma_{\rm eff}$.
         (c) The fitted frequency $\Omega$ vs $U$
             for $E_d^{f}/\Gamma_{\rm eff} =
             -E_d^{i}/\Gamma_{\rm eff} = 1$ (black
             circles), $3$ (red circles), and $10$ (green
             circles). Full lines display the analytical
             strong-coupling expression for $\Omega$,
             employing the numerical values of $V$. Arrows
             on the right-hand side mark the asymptotic
             $U \to \infty$ values of $\Omega$.
         (d) Same as (c) for the decay time $\tau$.
         All remaining parameters are the same as in
         Fig.~\ref{fig:Fig2}.
         }
\label{fig:Fig3}
\end{figure}

The extended time scales and supreme accuracy of
the hybrid NRG-DMRG allow for a detailed quantitative
analysis of the interaction-enhanced oscillations,
which was previously impossible using the hybrid
NRG-Chebyshev approach~\cite{Hybrid-NRG}.
The understanding of the interaction-enhanced
oscillations employs a strong-coupling expansion
in $1/U$. For $U \to \infty$, the impurity level
and zeroth Wilson shell decouple from the rest of the
chain, being confined to a combined valence of one.
Within this subspace, the two eigenstates of the
local Hamiltonian have the energies $\epsilon_{\pm}
= (E_d/2) \pm \sqrt{ (E_d/2)^2 + V^2 }$, hence
$n_d(t)$ displays quantum beats with the frequency
$\Omega = \epsilon_{+} - \epsilon_{-} =
2\sqrt{ (E_d/2)^2 + V^2 }$~\cite{Hybrid-NRG}. For
large but finite $U$, the coherent oscillations
are damped by the residual coupling to the rest
of the chain, which introduces a finite lifetime
of the state $\epsilon_{+}$. We expand to order
$1/U$ about the $U \to \infty$ limit, which yields
the residual coupling to the rest of the chain.
Using Fermi's golden rule, the decay time is
found to be
\begin{equation}
\tau^{-1} = \pi
            \left (
                    \frac{8 D}{\pi^2 U}
            \right )^2
            \frac{V^2}{\sqrt{ (E_d/2)^2 + V^2 } } .
\end{equation}
Further neglecting rearrangements of the bath
electrons (themselves being controlled by $1/U$),
we deduce the functional form
\begin{eqnarray}
n_d(t) &=& A \left [
                     e^{-t/2\tau} \cos(\Omega t)
                     \sqrt{1 - e^{-t/\tau}\!\cos^2 \theta}
                     - e^{-t/\tau}\!\sin \theta
             \right ]
\nonumber \\
       &+& n_{\rm eq} \left ( 1 - e^{-t/\tau} \right )
        + n_d(0) e^{-t/\tau} ,
\label{fit-function}
\end{eqnarray}
where $n_{\rm eq}$ is the equilibrated long-time
occupancy, while $A$ and $\theta$ have no direct
relation to any simple observable.

Panels (a) and (b) of Fig.~\ref{fig:Fig3} show typical
fits of $n_d(t)$ to Eq.~(\ref{fit-function}),
using $\tau$, $\Omega$, $A$, $\theta$, and
$n_d(0) = 1 - n_{\rm eq}$ as fitting parameters. The
fitting range, $t\cdot\Gamma_{\rm eff} \leq 10$, was
carefully chosen to exclude any discretization error.
Evidently, Eq.~(\ref{fit-function}) well describes
the numerical curves, further validating the
expansion in $1/U$. The extracted values of
$\Omega$ and $\tau$, plotted in panels (c) and (d)
for different quenches, practically coincide
above $U/D \approx 10$ with the analytical predictions,
confirming, in particular, that $\tau \propto U^2$.

\paragraph{Summary.}

A new hybrid NRG-DMRG approach was devised that
largely eliminates discretization errors hampering
the TD-NRG, while boosting the TD-DMRG to time
scales orders of magnitude beyond its natural reach.
The approach allows access to exceptionally long
times with unparallel accuracy, as demonstrated
by a detailed analysis of the interaction-enhanced
oscillations in the IRLM. These outstanding
capabilities open the door, so we hope, to accurate
investigations of systems and coupling regimes
that so far remained well beyond reach.

\paragraph{Acknowledgments}
This work was supported by the German-Israeli
Foundation through grant no.\ 1035-36.14, and
by the Deutsche Forschungsgemeinschaft under
AN 275/6-2 (F.G.\ and F.B.A).

%
\end{document}